\documentclass[aps,pra,reprint,showpacs,twocolumn,superscriptaddress,eqsecnum,floatfix]{revtex4-2}
\usepackage{placeins}
\usepackage{graphicx}
\usepackage{dcolumn}
\usepackage{bm}
\usepackage{amsmath}
\usepackage[utf8]{inputenc}
\usepackage{physics}
\usepackage{tabularx} 
\usepackage{array}
\usepackage{soul}
\usepackage{makecell}
\usepackage{multirow}
\usepackage{booktabs} 
\usepackage{tabularx}
\usepackage{mathrsfs}
\usepackage{xcolor}
\usepackage{color}
\usepackage{float}
\usepackage{appendix}
\usepackage{enumitem}
\usepackage{makecell}
\usepackage{tikz}
\usepackage{natbib,hyperref}
\usepackage{xr}
\hypersetup{colorlinks=true,linkcolor=blue,urlcolor=blue,citecolor=blue}

\newcommand{\RN}[1]{%
  \textup{\uppercase\expandafter{\romannumeral#1}}%
}

\begin{document}

\preprint{APS/123-QED}

\title{Classical reservoir approach for efficient molecular ground state preparation}

\author{Zekun He}
 \email{zh168@georgetown.edu}
 \affiliation{%
Department of Physics, Georgetown University, Washington DC 20057, USA
}%

\author{Dominika Zgid}
\email{zgid@umich.edu, dominika.zgid@fuw.edu.pl}
\affiliation{Department of Chemistry, University of Michigan, Ann Arbor, Michigan 48109, USA}
\affiliation{Department of Physics, University of Michigan, Ann Arbor, Michigan 48109, USA}
\affiliation{Faculty of Physics, University of Warsaw, 02-093 Warsaw, Poland}

\author{A.~F.~Kemper}
\email{akemper@ncsu.edu}
\affiliation{Department of Physics and Astronomy, North Carolina State University, Raleigh, North Carolina 27695, USA}

\author{J. K. Freericks}%
\email{james.freericks@georgetown.edu}
\affiliation{%
Department of Physics, Georgetown University, Washington DC 20057, USA
}%

\date{\today}

\begin{abstract}
Ground state preparation is a central application of quantum algorithms for electronic structure. We introduce the classical reservoir approach, a low-cost variational ansatz tailored to near-term hardware, requiring only nearest-neighbor interactions on a machine with square-lattice connectivity. Unlike traditional methods built from the classically efficient Hartree–Fock theory, our ansatz operates in localized molecular orbitals to study previously unexplored regions of the variational parameter space. Numerical benchmarks demonstrate chemical accuracy across diverse systems and bond lengths; notably, significantly reduced circuit depths are attainable when relaxed error thresholds (e.g., tens of $\text{m}E_h$) are permissible.
We benchmark the method on hydrogen chains, $\mathrm{N}_2$, $\mathrm{O}_2$, $\mathrm{CO}$, $\mathrm{BeH}_2$, and $\mathrm{H}_2\mathrm{O}$, the latter corresponding to an effective 24-qubit calculation.
\end{abstract}

\maketitle

\section{Introduction}Rapid progress in superconducting quantum hardware, including Google’s square lattice architectures~\cite{google2025quantum} and IBM’s heavy hex architectures~\cite{bravyi2024high}, is making it increasingly feasible to execute sophisticated algorithms for quantum many body problems~\cite{google2020hartree}. Problems whose Hilbert space dimension grows exponentially with system size $N$, such as ground state preparation, are particularly attractive targets. Quantum devices offer two key advantages: they bypass the severe memory constraints that limit classical simulation, and they provide a path to accurate treatment in strongly correlated regimes where classical methods degrade. For example, coupled cluster with perturbative triples (CCSD(T)) is widely regarded as the gold standard near equilibrium~\cite{valeev2008coupled} but loses reliability in a strongly correlated regime. More advanced approaches, such as auxiliary field quantum Monte Carlo (AFQMC)~\cite{sukurma2024toward,motta2018ab,zhang2003quantum,Unbiasing2025jiang}, can help, although they are affected by the fermionic sign problem and are sign free only in special symmetry cases~\cite{li2019sign,robust2023grossman}. These considerations motivate efficient quantum algorithms tailored to near term devices for addressing classically hard cases.

In the domain of quantum algorithms, many variational approaches have been proposed prior to this work~\cite{rapid2025berry,towards2017motta,
cerezo2021variational,motta2021low,motta2023bridging,
chen2021quantum,grimsley2019adaptive,matsuzawa2020jastrow,progress2022cao}, offering alternatives to other philosophies such as imaginary-time evolution~\cite{motta2020determining,gomes2021adaptive,mcardle2019variational} and subspace methods~\cite{motta2024subspace,mejuto2023quantum}. 
Focusing on near-term devices with limited coherence time, circuit depth, and qubit connectivity, we introduce a particularly simple variational algorithm tailored to square-lattice architectures, in which all two-qubit operations are strictly local two-body terms, eliminating the need for SWAP gates~\cite{kivlichan2018quantum}. 
These operations include same-spin hopping and opposite-spin correlation terms. 
The approach is motivated by a cooling perspective, with operators derived from the classical reservoir framework~\cite{he2025efficient}. 

To highlight the improvements introduced by our algorithm, we draw attention to two established directions that have proven effective. 
The first direction is the family of ansätze based on unitary coupled cluster singles and doubles (UCCSD)~\cite{chen2021quantum,chen2022low} and ADAPT-VQE~\cite{grimsley2019adaptive,nonvariational2025tang,traore2024shortcut}, both of which include double excitations in their wave functions.
In contrast, our work achieves comparable or better accuracy than the state-of-the-art results reported in Refs.~\cite{traore2024shortcut,chen2022low} with fewer CNOT gates. This efficiency stems from the classical reservoir approach, which avoids explicit double excitations—involving the creation and annihilation of two particles, which is much more costly than single excitations~\cite{yordanov2020efficient}— and also avoids long-range terms, which necessitate SWAP chains to map non-local interactions onto hardware with limited connectivity.

The second established direction relates to the Jastrow method~\cite{matsuzawa2020jastrow,neuscamman2013communication,haupt2023optimizing} and its restriction to local operations in the local unitary cluster Jastrow (LUCJ) ansatz~\cite{motta2023bridging}, which likewise avoids double excitations and employs only local operators. By adopting a different philosophy, our algorithm extends variational strategies to achieve competitive accuracy across both strongly and weakly correlated regimes while requiring fewer resources.

\begin{figure*}[htp]
    \begin{centering}
        \includegraphics[width=2\columnwidth]{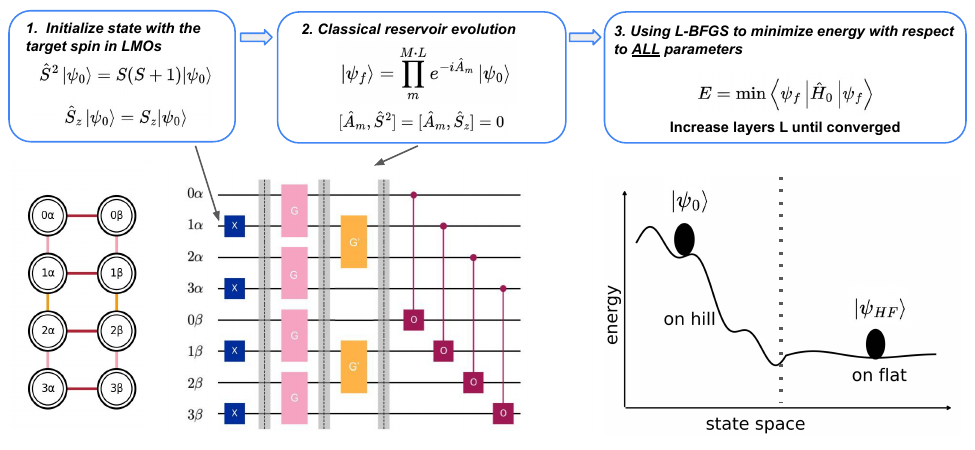}
        \caption{
Top: flow chart of the classical reservoir method.  
Bottom left: square qubit layout with the mapping of the spin orbitals, where $\alpha$ denotes spin-up electrons and $\beta$ denotes spin-down electrons.
Bottom center: quantum circuit diagram of the method, beginning with a single layer of $X$ gates to generate a typical high-energy initial state (i.e., a doubly occupied configuration) as shown in Eq.~\ref{eq:ini state}, followed by one ansatz layer consisting of two hopping layers implemented as Givens rotations (denoted by the symbol $G$) between adjacent same spin orbitals and one on-site potential layer between opposite spin orbitals implemented with $ZZ$ and $Z$ gates (denoted by the symbol $O$).  
Bottom right: schematic representation of the initialization, illustrating the idea of starting from a high-energy initial state rather than the Hartree–Fock state to ``roll the rock'' down to the lowest point, i.e., preparing the ground state.
}
        \label{fig:overview_fig}
    \end{centering}
\end{figure*}

\section{Formalism}The electronic structure Hamiltonian, whose ground state we aim to prepare, can be expressed in second quantization as
\begin{equation}
H = \sum_{p,q} h_{pq}\, c_p^\dagger c_q
  + \frac{1}{2} \sum_{p,q,r,s} h_{pqrs}\, c_p^\dagger c_q^\dagger c_r c_s ,
\label{eq:hamiltonian}
\end{equation}
where $h_{pq}$ are one-electron integrals, $h_{pqrs}$ are two-electron Coulomb
integrals, and $c_p^\dagger, c_q$ are fermionic creation and annihilation
operators acting onto spin orbitals.

In traditional quantum chemistry calculations, Hartree--Fock (HF) theory~\cite{mcardle2020quantum} is used as an initial calculation designed to provide starting orbitals. Subsequently, such a calculation is followed by a correlated calculation, such as coupled clusters (CC) for example, CCSD with greatly simplified amplitude equations.  Even in the quantum hardware era, the variational unitary extension of CC, UCCSD, remains in the same spirit. 
It is initialized from a HF reference state (the best mean field approximation state), and the initial guess for optimizing the single and double excitations is based on either CCSD amplitudes or 
second-order Møller--Plesset (MP2) perturbation 
theory~\cite{hirsbrunner2024beyond,chen2021quantum}. More recently, to 
avoid the high cost of double excitations and long range interactions, the LUCJ ansatz~\cite{motta2023bridging} was introduced. By employing a more sophisticated double-factorization 
method~\cite{motta2021low},  double excitations such as the second term in Eq.~\ref{eq:hamiltonian} can be translated into the LUCJ ansatz expression.

Yet, it is unclear whether these classically efficient methods provide optimal starting points when variational quantum circuits are the workhorse. Moreover, CCSD is known to underperform in a strongly correlated regime~\cite{bulik2015can} and could reasonably be expected to provide a flawed starting point. Most quantum algorithms, in the absence of widely adopted alternatives that genuinely depart from this paradigm,  still initialize from an HF state and employ CCSD- or MP2-based heuristics to construct a better initial guess. This strategy is preferable to an uninformed random choice, but its optimality remains unknown.

In this work, we step outside this framework. We introduce a cooling-inspired quantum algorithm, as illustrated in Fig.~\ref{fig:overview_fig}, and show numerically that it can discover more efficient preparation paths without relying on classical heuristics. We replace HF spin orbitals with localized molecular orbitals (LMOs) 
constructed by the Edmiston and Ruedenberg procedure~\cite{Edmiston1963localized}, 
which maximizes the electronic self-repulsion energy, resulting in spatially localized 
spin orbitals. 
The algorithm then starts from a simple product state that is an 
eigenstate of the total spin operator $\hat{S}^2$ rather than the HF state, and applies 
classical reservoir operators~\cite{he2025efficient} as a compact unitary. 
The associated parameter amplitudes are numerically optimized by gradient descent to drive the 
energy toward the ground state. 

This design is motivated by both physical intuition and practical considerations for the quantum hardware. For the choice of orbitals, LMOs concentrate electronic correlation within each spin orbital, so modifying double occupancy produces a stronger and more targeted effect on the quantum state than when HF orbitals are used, making this ansatz more effective per parameter. Moreover, LMOs can substantially reduce the $L_1$ norm of the electronic Hamiltonian relative to HF orbitals (by as much as 76\% for larger molecules such as $\mathrm{HNC_7H_{14}}$) thereby directly lowering measurement costs in quantum circuits~\cite{Koridon2021Orbital,berry2020time}.

\begin{equation}
\left|\psi_0\right\rangle =
\left|
\underbrace{00}_{\phi^{\mathrm{LMO}}_{1}}
\overbrace{\uparrow \downarrow}^{\phi^{\mathrm{LMO}}_{2}}
\underbrace{00}_{\phi^{\mathrm{LMO}}_{3}}
\overbrace{\uparrow \downarrow}^{\phi^{\mathrm{LMO}}_{4}}
\ldots
\underbrace{00}_{\phi^{\mathrm{LMO}}_{N-1}}
\overbrace{\uparrow \downarrow}^{\phi^{\mathrm{LMO}}_{N}}
\right\rangle
\label{eq:ini state}
\end{equation}

For the choice of the initial state, because the method is cooling based, such a state need not be low energy; we require only that it is trivial to prepare. In practice, such readily prepared states are often high in energy, which can be advantageous. A total-spin eigenstate, composed primarily of doubly occupied spin orbitals, as in Eq.~\ref{eq:ini state}, provides such an ideal starting point, exhibiting larger gradient norms along multiple descent directions and enabling broader exploration of parameter space (supporting diverse initial parameter guesses that easily descend in energy).
A schematic of such a rapid energy descent is shown in the bottom-right panel of Fig.~\ref{fig:overview_fig}. In practice, the energy can be quickly reduced from a high energy starting  configuration to the HF reference level within only a few optimization iterations, indicating that little is lost by not starting from the HF state.

By contrast, initializing from the HF state often restricts viable starting amplitudes: one must either rely on informed guesses or set them to zero, as in ADAPT-VQE~\cite{grimsley2019adaptive}, since most other choices raise the energy and undermine the intended advantage of the HF reference. Details of the initial parameter choices used in this work are provided in the Appendix. We conjecture that this flexibility to explore a larger region of parameter space, previously inaccessible when constrained by HF initialization, contributes to the improved resource efficiency observed here.

To preserve the correct total spin, the subsequent evolution must be spin-conserving. Although UHF can improve mean-field descriptions in strongly correlated regimes by breaking spin symmetry and thereby mimicking static correlation~\cite{hollett2011two}, the FCI energy remains identical regardless of whether RHF or UHF orbitals are used. Consequently, when the goal is to approach the FCI energy, we adopt a spin-conserving ansatz with restricted spin orbitals to ensure that the ground state remains an eigenstate of \(\hat{S}^2\) without any loss of accuracy.  Enforcing spin conservation provides several practical advantages:  
(i) it reduces the optimization parameter space by linking amplitudes in the spin-up and spin-down sectors;  
(ii) it prevents spin contamination when aiming for chemical accuracy; and  
(iii) it can, in some cases, widen the energy gap compared to non–spin-conserving settings.

\section{Method}Following Ref.~\cite{he2025efficient}, the classical reservoir operators are partitioned into three commuting groups. For an even number of spatial orbitals, the first and second groups contain nearest neighbor single excitations arranged 
according to the Jordan--Wigner fermionic encoding with interleaved orbital pairs. 
The first group consists of hopping terms acting on the pairs 
$\langle 1,2 \rangle, \langle 3,4 \rangle, \dots, \langle N-1, N \rangle$. 
The second group contains the interleaved pairs 
$\langle 2,3 \rangle, \langle 4,5 \rangle, \dots, \langle N-2, N-1 \rangle$, 
with interaction strengths parameterized by $\vec{\lambda}$. 
The final group includes the double occupancy number operators acting on all spatial orbitals, 
with coefficients parameterized by $\vec{\lambda}'$.

We denote the first hopping set by $\hat{T}(\vec{\lambda})$, 
the second by $\hat{T}'(\vec{\lambda})$, 
and the double occupancy operators by $\hat{U}(\vec{\lambda}')$.With these definitions, the classical reservoir ansatz can be written as  
\begin{align}
\left|\psi_f\right\rangle &= \prod_{l=1}^{L} e^{-i \hat{U}_l(\vec{\lambda}^{\prime})}
                             e^{-i \hat{T}^{\prime}_l(\vec{\lambda})}
                             e^{-i \hat{T}_l(\vec{\lambda})}
                             \left|\psi_0\right\rangle ,\\
&= \prod_{m=1}^{M \cdot L} e^{-i \hat{A}_m} \left|\psi_0\right\rangle,
\label{eq:final_state}
\end{align}
where \(l\) indexes the layers. For a total ansatz depth \(L,\) each layer contains \(M = 2N - 1\) amplitude parameters, with \(N\) denoting the number of spatial orbitals. The cost function is the final evolved energy \(E = \langle \psi_f | H | \psi_f \rangle\), and gradient descent is employed to optimize the amplitudes for the \((2N-1)L\) parameters. A flow chart of the full algorithm is provided in Fig.~\ref{fig:overview_fig}.


\section{Numerical results}In the main text, we present results for $\mathrm{H_2O}$, $\mathrm{N}_2$ and hydrogen chains which are three widely popular benchmark systems. Additional cases such as $\mathrm{CO}$ and $\mathrm{BeH}_2$ are summarized in the Appendix. All systems studied here exhibit qualitatively similar behavior, consistently achieving chemical accuracy.

\begin{figure}
\begin{centering}
    \includegraphics[width=0.9\columnwidth]{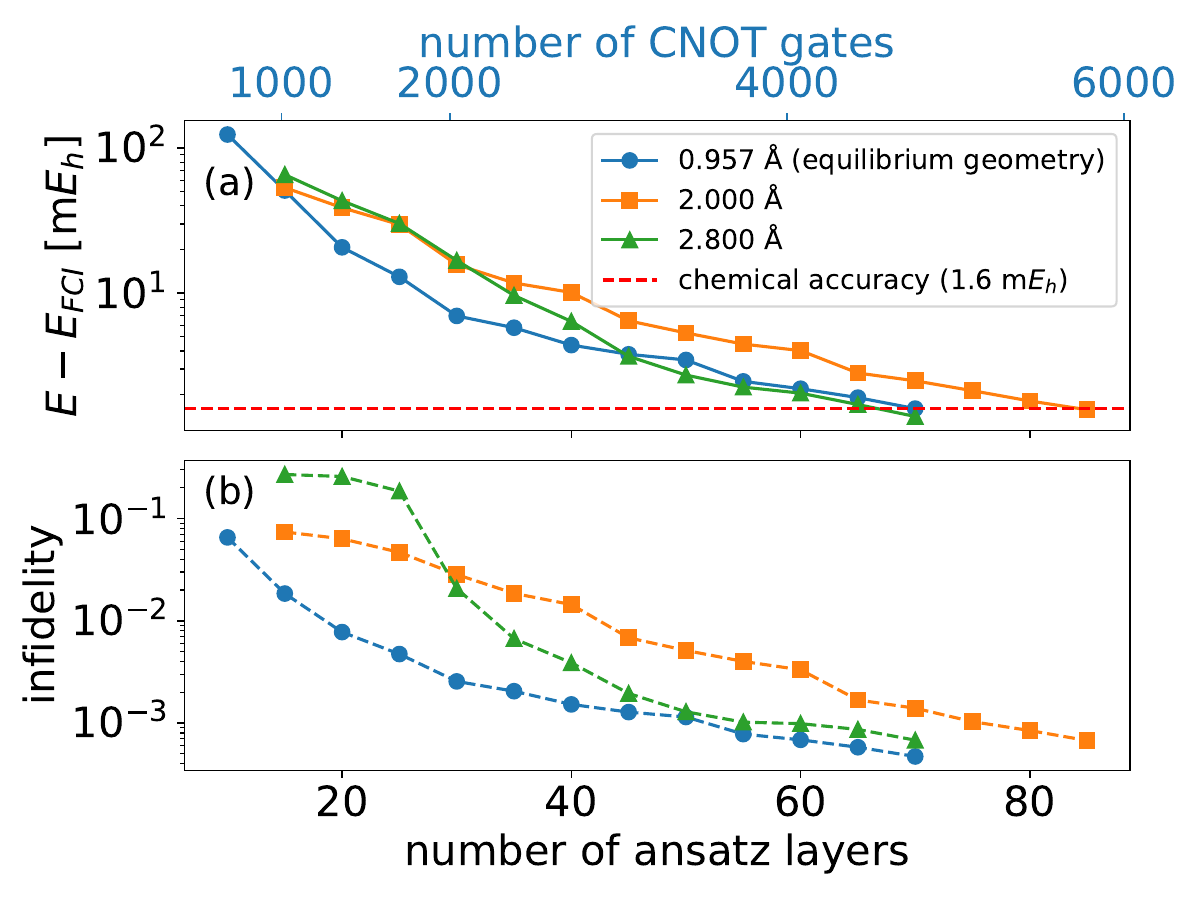}
\end{centering}
\caption{
(a) Energy difference from FCI for $\mathrm{H_2O}$ using the 6-31G atomic basis at various geometries, plotted as a function of the number of ansatz layers. A secondary (top) $x$ axis shows the corresponding CNOT gate count.  
The red dashed line indicates the chemical-accuracy threshold.  
(b) Infidelity for the same set of geometries as a function of the number of ansatz layers.
}
\label{fig:water_energy_infidelity}
\end{figure}

\begin{figure}
\begin{centering}
    \includegraphics[width=0.9\columnwidth]{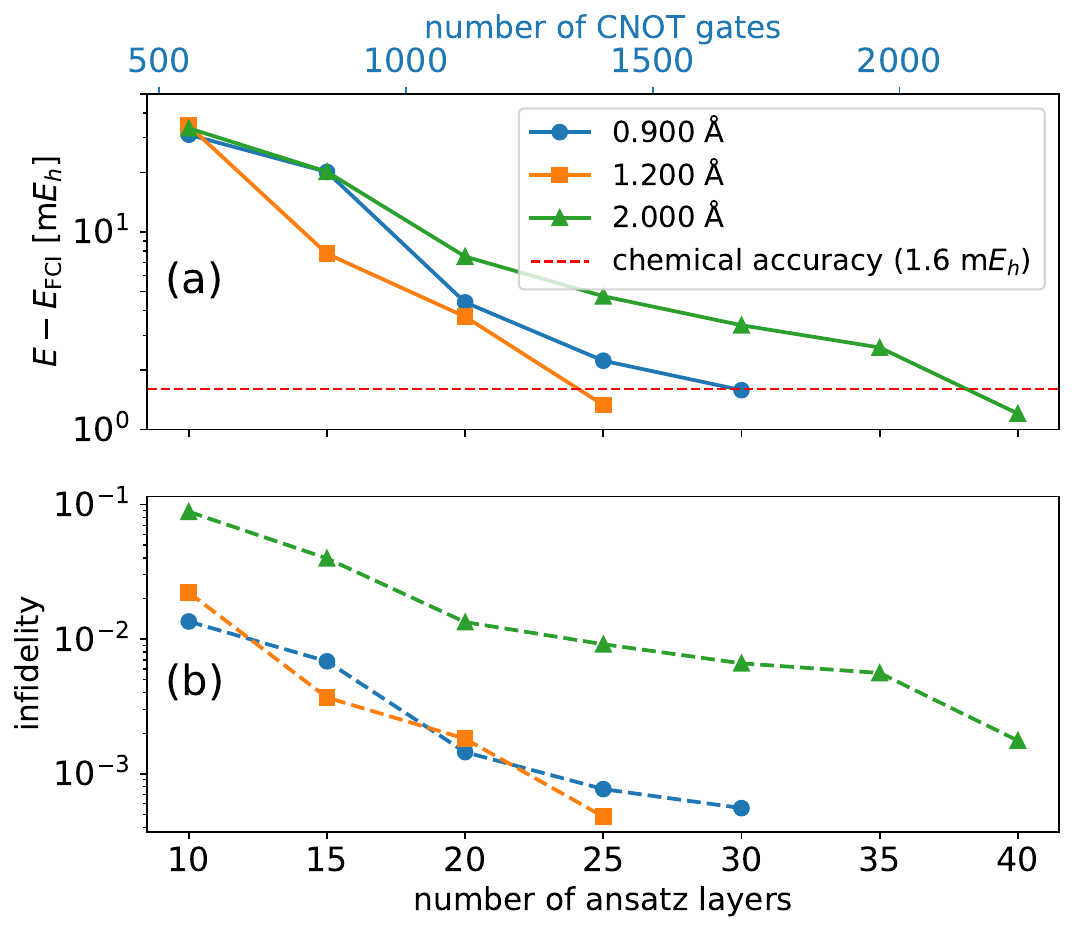}
\end{centering}
\caption{
(a) Energy difference from FCI for $\mathrm{N_2}$ using the STO-6G atomic basis at various geometries, plotted as a function of the number of ansatz layers.
(b) Infidelity for the same set of geometries as a function of the number of ansatz layers.
}
\label{fig:energy_analysis_panels_NN}
\end{figure}

In Fig.~\ref{fig:water_energy_infidelity}, we present results for $\mathrm{H_2O}$ using the 6-31G atomic basis with a standard frozen-core approximation that freezes the oxygen $1s$ orbital, while in Fig.~\ref{fig:energy_analysis_panels_NN} we show results for $\mathrm{N}_2$ with the STO-6G basis. In both cases, the bond length is stretched symmetrically while keeping the bond angle fixed. From these data, we identify two key advantages of the method.

First, the algorithm prepares the ground state across a wide range of geometries, from weakly to strongly correlated, without requiring a substantial increase in the ansatz depth when moving from weak to strong correlation. As shown in Fig.~\ref{fig:water_energy_infidelity}, at \(r = 2.8~\text{\AA}\) the computational cost remains comparable to that at the equilibrium geometry. This behavior contrasts with many previous state-of-the-art approaches. For example, in Ref.~\cite{chen2021quantum}, a low-rank UCCSD ansatz for the same molecule and basis shows the energy error increasing from below \(1.6~\text{m}E_h\) to about \(20~\text{m}E_h\) as the bond becomes elongated.

Second, for the same $\mathrm{H_2O}$ system studied where ADAPT-VQE~\cite{grimsley2019adaptive} can also achieve chemical accuracy, the present method requires substantially fewer quantum resources. To estimate the two-qubit gate count, note that each ansatz layer consists of \(N-1\) hopping terms in each spin sector and \(N\) double-occupancy number operators, for a total of \(3N-2\) operators per layer. In the latest IBM qiskit gate decomposition~\cite{qiskit2024}, assigning two CNOT gates to each Givens rotation (an \(XX{+}YY\) gate) for every hopping term and also two CNOT gates to each \(ZZ\) operator for the number operators yields \(2(3N-2)=6N-4\) CNOT gates per layer. For \(N=12\), in our algorithm a depth of \(L=70\) is sufficient to reach chemical accuracy. This corresponds to 4{,}760 CNOT gates on a square-lattice quantum processor, which is available as the latest Google’s Willow chip~\cite{google2025quantum}. In comparison, prior state-of-the-art results using ADAPT-VQE with a qubit-excitation-based operator pool require 12{,}657 total CNOT gates on an all-to-all connectivity machine~\cite{traore2024shortcut}. In practice, this resource gap can be even larger since many ADAPT-VQE operators correspond to long-range interactions that necessitate additional SWAP gates~\cite{kivlichan2018quantum} on hardware with limited connectivity.

It is important to recognize both the similarity to—and, more importantly, the distinction from—the unitary cluster Jastrow (UCJ) ansatz, defined as $\ket{\psi_{\mathrm{UCJ}}} = \prod_{\mu=1}^{L} e^{\hat{K}_\mu} e^{i\hat{J}_\mu} e^{-\hat{K}_\mu} \ket{\psi_{\mathrm{HF}}}$, where $\hat{K}_\mu = \sum_{p q,\sigma} K^{\mu}_{p q}\, \hat{c}^{\dagger}_{p\sigma} \hat{c}_{q\sigma}$ and $\hat{J}_\mu = \sum_{p q,\sigma \tau} J^{\mu}_{p q,\sigma \tau}\, \hat{n}_{p\sigma} \hat{n}_{q\tau}$. The operators $\hat{K}_\mu$ and $\hat{J}_\mu$ can be chosen with additional constraints—such as enforcing spin conservation or imposing locality consistent with the connectivity of the target quantum hardware—resulting in the LUCJ ansatz~\cite{motta2023bridging}.

Two key features distinguish this approach from UCJ and lead to different behavior.\\
\textbf{(i) Operator content and locality.} The kinetic-like term in this work is restricted to nearest-neighbor hopping operators in both the qubit layout and the spin-orbital ordering, similar to LUCJ~\cite{motta2023bridging} but not to the general UCJ ansatz. By the closure property of single-excitation commutators, the Lie algebra generated by local single excitations closes within the span of nearest-neighbor hoppings, so these operators already form a complete basis. One could, in principle, extend the pool by introducing current operators; in UCJ this arises by allowing general complex amplitudes in $\hat{K}_\mu$. However, prior work~\cite{matsuzawa2020jastrow} showed that restricting $\hat{K}_\mu$ to real amplitudes yields only current operators, which do not lower the energy below the reference and are less effective in practice. Our numerical results corroborate this, so we retain only hopping terms. For the Jastrow-like sector, our number operators are strictly on site (double occupancy within the same spatial orbital). Unlike UCJ, we do not include inter-orbital correlations between all pairs of spin orbitals; and relative to LUCJ, we also omit same-spin correlations to further reduce parameters we find to be less impactful.

\textbf{(ii) Layer structure.} UCJ/LUCJ adopt the sandwich form $e^{\hat{K}_\mu} e^{i\hat{J}_\mu} e^{-\hat{K}_\mu}$, where the kinetic piece acts mainly as a basis rotation, implementing an effective double-factorization~\cite{motta2021low}. In our design, the final $e^{-\hat{K}_\mu}$ is unnecessary: we interleave hopping layers directly with an on-site potential layer. At a fixed parameter budget this removes gates and lowers the circuit depth.

\begin{table}
\centering
\small 
\setlength{\tabcolsep}{6pt}  
\renewcommand{\arraystretch}{1.15} 
\begin{tabular}{@{}l rr rr@{}} 
\toprule
\multirow{2}{*}{\textbf{system}}
& \multicolumn{2}{c}{\textbf{UCJ}}
& \multicolumn{2}{c}{\textbf{this work}} \\
\cmidrule(lr){2-3}\cmidrule(l){4-5}
& \textbf{params} & $\boldsymbol{\Delta E}$ (mE\!h)
& \textbf{params} & $\boldsymbol{\Delta E}$ (mE\!h) \\
\midrule
\multirow{2}{*}{$\mathrm{H}_8$}
& 135 & 12   & 135 & 6.3 \\
& 270 & 1    & 225 & 0.5 \\
\midrule
\multirow{2}{*}{$\mathrm{H}_{10}$}
& 209 & 18   & 209 & 6 \\
& N/A & N/A  & 399 & 0.7 \\
\bottomrule
\end{tabular}
\caption{Comparison of parameter counts and energy errors between UCJ~\cite{matsuzawa2020jastrow} and the present work for 8- and 10-hydrogen chains at $r = 2~\text{\AA}$ (moderately strongly correlated).}
\label{tab:kucj_vs_thiswork}
\end{table}

These simplifications matter: with a fixed total number of parameters, eliminating less effective current operators and long-range density terms lets us allocate more parameters to the more impactful hopping and on-site interactions. As shown in Tab.~\ref{tab:kucj_vs_thiswork} for hydrogen chains, the original UCJ ansatz, which includes long-range kinetic and Jastrow interactions, underperforms the strictly local interactions used here. This shows that enforcing locality reduces circuit depth and complexity without sacrificing accuracy—and can even improve it. We observe similar behavior with IBM’s software package~\cite{qiskit2024} for other molecules such as $\mathrm{N_2}$ and $\mathrm{CO}$, supporting the conclusion that well-designed local operators are both sufficient and efficient in our framework.

Finally, we examine the algorithm under a more restrictive CNOT budget, limiting circuits to approximately \(1000\) CNOT gates in order to evaluate its performance under realistic near-term resource constraints. In the Appendix, we present \(\mathrm{H_2O}\) as a single illustrative example; however, the conclusion is general: across all molecules studied, an ansatz with about \(N\) layers achieves \(\approx 0.99\) fidelity at the equilibrium geometry. When this equilibrium solution is annealed into the strongly correlated region, it maintains high fidelity. These results indicate that the method provides substantial ground state overlap even at low CNOT budgets, yielding an excellent initial state for quantum phase estimation on future fault tolerant quantum hardware, where chemical accuracy is expected without additional variational optimization~\cite{kitaev2002classical}.

\section{Conclusion}This work shows that ground states can be prepared with chemical accuracy at substantially lower computational cost, making the method well suited to near term quantum hardware. More broadly, it advances a quantum first design philosophy for algorithms: they need not be constrained by classical heuristics or efficiencies. We hope this perspective motivates further quantum native designs for preparing ground states of many body quantum systems.




\section{Acknowledgments}This work was supported by the Department of Energy, Office of Basic Energy Sciences, Division of Materials Sciences and Engineering under grant no. DE-SC0023231. J.K.F. was also supported by the McDevitt bequest at Georgetown. D.Z was supported by the National Science Foundation under grant number CHEM-2154672.

\section{Data Availability}The data that support the findings of this article as well as the python code that run the calculations are openly available at~\cite{data}.



\bibliography{bib.bib}


\clearpage
\section{Appendix A — CNOT gate budget case study}

\begin{figure}
\begin{centering}
    \includegraphics[width=0.9\columnwidth]{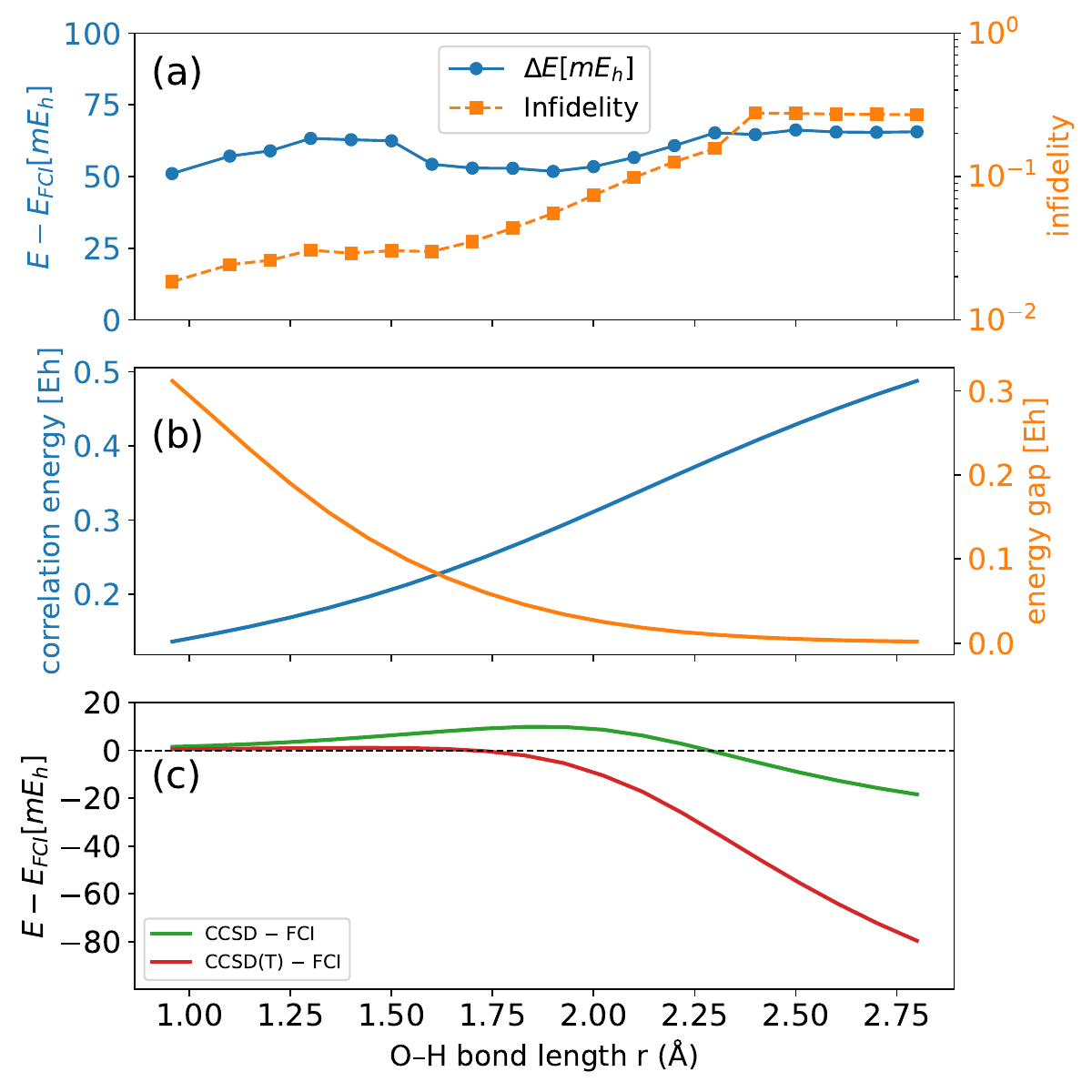}
\end{centering}
\caption{
(Color online) (a) Energy difference from FCI for $\mathrm{H_2O}$ using 15 ansatz layers as a function of the O–H bond length.  
(b) Correlation energy and the gap between the ground state and the first excited state within the same total spin sector.  
(c) Energy difference from FCI for $\mathrm{H_2O}$ computed using CCSD and CCSD(T) from PySCF~\cite{sun2020recent} as a function of the O–H bond length.
}
\label{fig:energy_analysis_panels_water}
\end{figure}

In Fig.~\ref{fig:energy_analysis_panels_water} we present a detailed study where the number of ansatz layers used ($15$) is close to the number of spatial orbitals ($12$), in order to investigate performance in this near matched setting. For $\mathrm{H_2O}$ with this ansatz depth, the total CNOT gate count is $1020$. We use this fixed ansatz depth to sweep the O–H bond lengths while keeping the bond angle fixed at its equilibrium value of $104.5^\circ$. The annealing procedure from the equilibrium geometry to the more strongly correlated geometries consists of two consecutive sweeps across geometries. 

First, a forward sweep is carried out starting from the equilibrium geometry and proceeding to progressively longer bond lengths. Each optimization is initialized with the converged parameters from the previous, shorter bond length geometry. Once the forward sweep is complete, a reverse sweep is performed from the longest bond length back to the shortest. During this second pass, if the reverse sweep produces a lower energy at any geometry, that result replaces the corresponding forward pass value; otherwise, the original result is retained. 

This two sweep annealing procedure stabilizes the optimized energy across the entire potential energy curve and reduces optimization variance, leading to a smoother and more reliable energy profile along the bond scan. Panel~(a) shows that the infidelity increases smoothly from about $10^{-2}$ at equilibrium to about $10^{-1}$ in the strongly correlated regime, while the energy error remains around $50\,\mathrm{mE_h}$. This deviation is small compared with both the energy gap in the same spin sector and the total correlation energy, as illustrated in panel~(b). More importantly, this annealing protocol maintains substantial overlap with the ground state even in the strongly correlated region, where the energy gap narrows to about $1.7\,\mathrm{mE_h}$ at a bond length of $2.8\,\text{\AA}$. Despite such a small gap, an energy error of roughly $50\,\mathrm{mE_h}$ still corresponds to about $0.7$ fidelity, indicating that the annealing approach avoids collapse to an orthogonal excited state. Panel~(c) compares classical coupled cluster benchmarks, showing that CCSD and CCSD(T) become unreliable in the strongly correlated regime; for example, at a bond length of $2.8\,\text{\AA}$ the CCSD(T) error reaches about $80\,\mathrm{mE_h}$.

\section{Appendix B — Summary of data}  

All molecules studied in this work, together with the quantum resources required to reach chemical accuracy, are summarized in the following table.

Table~\ref{tab:molecule_comparison} reports, for each molecule, the number of ansatz layers and the corresponding CNOT gate count needed to achieve chemical accuracy. These quantities provide a concise measure of the quantum resources required by the proposed method for systems of increasing complexity, with a selection designed to span a representative range of commonly used benchmark molecules.

\begin{table*}
\centering
\renewcommand{\arraystretch}{1.3}
\setlength{\tabcolsep}{8pt}
\begin{tabular}{|l|c|c|c|c|c|}
\hline
\textbf{molecule (basis)} &
\shortstack{\textbf{Hamiltonian}\\\textbf{dimension}} &
\shortstack{\textbf{bond length}\\(\AA)} &
$\boldsymbol{\Delta E}$ (m$E_\mathrm{h}$) &
\shortstack{\textbf{ansatz}\\\textbf{layer count}} &
\shortstack{\textbf{CNOT}\\\textbf{gate count}} \\
\hline
H$_2$O (6-31G)      & 245{,}025 &   0.957 (equilibrium)          &     1.6        &    70         &    4{,}760        \\ \hline
H$_2$O (6-31G)      & 245{,}025 &    2.000         &      1.6       &     85        &      5{,}780      \\ \hline
H$_2$O (6-31G)      & 245{,}025 &    2.800         &      1.4       &      70       &    4{,}760        \\ \hline
N$_2$ (STO-6G)      &    14{,}400        &    0.900         &    1.6         &      30       &      1{,}680      \\ \hline
N$_2$ (STO-6G)      &    14{,}400        &     1.200(equilibrium)         &   1.3          &      25       &   1{,}400          \\ \hline
N$_2$ (STO-6G)      &      14{,}400      &       2.000      &    1.2         &    40         &      2{,}240      \\ \hline
$\mathrm{H}_{10}$ (STO-6G) &  63{,}504 &     1.000        &     0.7        &   40          &    2{,}240        \\ \hline
$\mathrm{H}_{10}$ (STO-6G) &  63{,}504 &     2.000       &             0.7&             21&            1{,}176\\ \hline
$\mathrm{H}_{8}$ (STO-6G)  & 4{,}900  &      1.000       &    0.5         &     25        &      1100      \\ \hline
$\mathrm{H}_{8}$ (STO-6G)  & 4{,}900  &      2.000       &             0.5&             15&            660\\ \hline
$\mathrm{BeH_2}$ (6-31G)   & 81{,}796  &  1.326(equilibrium)         &    1.2         &    30         &         2{,}220   \\ \hline
$\mathrm{CO}$ (STO-6G)     & 14{,}400  &   1.128(equilibrium)            &       1.4      &    35         & 1{,}960           \\ \hline
$\mathrm{O_2}$ (STO-6G)    & 1{,}200  &  1.210(equilibrium)           &     1.5        &    20         &  1{,}120          \\ \hline
\end{tabular}
\caption{Summary of the molecules studied in this work. Equilibrium geometries are accurate at the level of the atomic basis sets used. All cases are studied with a restricted spin basis, except for $\mathrm{O}_2$, which is treated as a restricted open shell system and serves as a contrasting setting.}
\label{tab:molecule_comparison}
\end{table*}

\section{Appendix C — Optimization details}

All calculations in this work use L-BFGS as implemented in PyTorch on GPU hardware~\cite{paszke2017automatic}. We set the convergence criterion when either the absolute change in energy between successive iterations falls below $10^{-8}\,E_\mathrm{h}$ or the Euclidean norm of the gradient is smaller than $10^{-4}$. The maximum number of iterations is set to $5000$.

A key finding of this work is that initialization of the ansatz parameters admits much greater freedom than is often assumed. We describe two practical strategies that we found to be effective and feasible for achieving chemical accuracy.

\paragraph{For chemical molecules}
\begin{enumerate}
\item \textbf{Equilibrium geometry, shallow depth.}
Work at the equilibrium geometry with a modest layer count (we use $L_0=5$). Draw uniform random initial parameters independently from several symmetric ranges,
\[
  [-\pi,\pi],\quad \bigl[-\tfrac{\pi}{2},\tfrac{\pi}{2}\bigr],\quad \bigl[-\tfrac{\pi}{4},\tfrac{\pi}{4}\bigr],\quad \bigl[-\tfrac{\pi}{8},\tfrac{\pi}{8}\bigr],
\]
and run $20$ randomized starts per range, for a total of $80$ trials. Select the lowest-energy result as the solution at $L_0$. One may also use variance-based or other statistical criteria to define a quantitative convergence threshold and determine how many trials are sufficient; in this work, however, we fixed the total number of trials to conserve computational resources, regardless of the number of spatial orbitals.

  \item \textbf{Increase depth with gentle noise.}
  To increase the layer count from $L$ to $L+\Delta L$ (for example, from $5$ to $10$, which is the typical step in this work), keep the first $L$ layers from the best solution and initialize the additional parameters with small uniform noise drawn from $[-0.01,0.01]$, $[-0.1,0.1]$, or $[-1,1]$. Choose the option that yields the lower post optimization energy. The small noise preserves the quality of the seed while providing enough exploration to escape shallow local minima.
  
\item \textbf{Anneal across geometries.}

Once a solution is obtained at the equilibrium geometry for a given ansatz layer depth, obtain results for other geometries as follows. First perform a forward sweep from the equilibrium geometry to progressively longer bond lengths (for stretching; for squeezing, use the opposite order), initializing each optimization with the converged parameters from the previous geometry. After the forward sweep is complete, perform a reverse sweep from long to short bond lengths. At each geometry, keep the reverse result only if it achieves a lower energy; otherwise retain the forward result. These two sweeps complete the calculation across all geometries at the specified layer count.

\end{enumerate}

\paragraph{For hydrogen chains}
\begin{enumerate}
  \item \textbf{Constant seeds across all parameters.}
  For any geometry and any layer count, initialize every parameter to the same constant $\theta_0$ and try the set
  \[
    \theta_0 \in \Bigl\{\pi,\ \tfrac{\pi}{2},\ \tfrac{\pi}{4},\ \tfrac{\pi}{8},\ \tfrac{\pi}{16},\ \tfrac{\pi}{32},\ \tfrac{\pi}{64}\Bigr\}.
  \]
  Run a short optimization budget (100 LBFGS iterations in this work) for each choice and keep the best performing seed. Then continue from that seed until convergence.

  \item \textbf{Depth and geometry sweeps.}
  Apply the selected constant seed approach independently at each depth and geometry. Empirically this converges faster than warm starting from a shallower depth for hydrogen chains.
\end{enumerate}


\end{document}